\documentclass[
nofootinbib,
twocolumn,
showpacs,
superscriptaddress,
prb,
aps
]{revtex4}
\usepackage{graphicx}
\usepackage{dcolumn}
\usepackage{bm}
\usepackage{color}
\def\BiR{Bi$_2$Sr$_{2-x}R_x$CuO$_y$}
\def\La-OP{La-Bi2201-OP}
\def\Eu-OP{Eu-Bi2201-OP}
\def\Tc{$T_{\mbox{\scriptsize{c}}}$}
\def\TMF{$T_{\mbox{\scriptsize{c}}}^{\mbox{\scriptsize{MF}}}$}
\def\Tonset{$T_{\mbox{\scriptsize{onset}}}$}
\def\T*{$T^*$}

\def\EF{$E_{F}$}
\def\kF{$k_{F}$}

\begin{document}
    \preprint{APS/12m3-QED}
    \title{
    Three energy scales characterizing the competing pseudogap state, the incoherent, and the coherent superconducting state 
    in high-\Tc\ cuprates 
    }

\author{Y. Okada}
    \affiliation{Department of Crystalline Materials Science, Nagoya University, Nagoya 464-8603, Japan}
\author{T. Kawaguchi}
    \affiliation{Department of Crystalline Materials Science, Nagoya University, Nagoya 464-8603, Japan}
\author{M. Ohkawa}
    \affiliation{Institute for Solid State Physics (ISSP), The University of Tokyo, Kashiwa 277-8581, Japan}
\author{K. Ishizaka}
    \affiliation{Institute for Solid State Physics (ISSP), The University of Tokyo, Kashiwa 277-8581, Japan}
\author{T. Takeuchi}
    \affiliation{EcoTopia Science Institute, Nagoya University, Nagoya 464-8603, Japan}
\author{S. Shin}
    \affiliation{Institute for Solid State Physics (ISSP), The University of Tokyo, Kashiwa 277-8581, Japan}
\author{H. Ikuta}
    \affiliation{Department of Crystalline Materials Science, Nagoya University, Nagoya 464-8603, Japan}


\begin{abstract}
We have studied the momentum dependence of the energy gap of Bi$_2$(Sr,$R$)$_2$CuO$_y$ by angle-resolved photoemission spectroscopy (ARPES),
particularly focusing on the difference between $R$=La and Eu.
By comparing the gap function and characteristic temperatures
between the two sets of samples, 
we show that there exist three distinct energy scales, $\Delta_{pg}$, $\Delta_{sc0}$, and  $\Delta_{sc0}^{eff}$, 
which correspond to \T*\ (pseudogap temperature), 
\Tonset\ (onset temperature of fluctuating superconductivity), and 
\Tc\ (critical temperature of coherent superconductivity).
The results not only support the existence of a pseudogap state below \T*\ that competes with superconductivity
but also the duality of competition and superconducting fluctuation at momenta around the antinode below \Tonset.
\end{abstract}
\pacs{74.72.-h, 74.72.Kf, 74.62.-c, 74.62.Dh}
\maketitle
\section{\label{sec:level1}Introduction}
    One of the significant differences between high-\Tc\ cuprates and conventional superconductors is the
    presence in the former of a pseudogap state above \Tc.
    Whether the pseudogap state is a precursor to superconductivity 
    or a state that competes with it 
    has been a matter of long-standing debate \cite{Timsk, Norman, Huffner, Emery, Varma, Chacravati}.
    To address this problem, 
    angle-resolved photoemission spectroscopy (ARPES) 
    is one of the most powerful techniques 
    since the momentum dependence of the energy gap is directly linked to the pseudogap issue.
    Many ARPES experiments have investigated this problem, 
    however, the data and their interpretations are still controversial.

    If the momentum dependence of the gap function is constituted by only one component, 
    the pseudogap state can be regarded as a precursor to the superconducting state.
    This picture has been supported 
    by some of the ARPES experiments, 
    which concluded that 
    the energy gap has predominantly a $d$-wave symmetry \cite{Kanigel,Valla, Wei, Meng, Nakayama, Shi}.
    An intimate relation between pseudogap and superconductivity 
    has been suggested also by high-frequency conductivity measurements \cite{Corson}, 
    the enhanced Nernst signal \cite{Xu}, 
    enhanced diamagnetism \cite{Wang_Nernst_diamagnetism}, 
    and the observation of quasi-particle interference pattern suggesting phase incoherent pairing gap above \Tc\ \cite{Lee_STM}.
    On the other hand, 
    other ARPES experiments suggested the existence of two gap components
    that depend differently on momentum, temperature, and carrier doping
    \cite{Tanaka,Kondo_PRL,Terashima,Lee, Vidya, Kondo_Nature,Hashimoto_Bi2201,He,Yoshida}.
    If the gap function consists of two components, 
    the presence of an additional order other than the $d$-wave superconductivity 
    must be assumed.
    This picture was supported further 
    by a recent experiment providing evidence 
    for the existence of a density wave state in high-\Tc\ cuprates 
    \cite{Doug,Hashimoto_Nat_Phys}.
    Here, if the co-existing state competes with superconductivity and suppresses \Tc\ 
    by reducing the number of paired electrons, 
    the superconducting order would significantly fluctuate
    as has been suggested \cite{Emery}.
    In this case, 
    both $competition$ and $superconducting$ $fluctuation$ should be consistently accounted for above \Tc.
    
    It is known that \Tc\ can be controlled both by the element $R$ and $x$ 
    in \BiR\ ($R$=rare earth elements) \cite{Nameki,Eisaki,Fujita,Hashimoto_Bi2201}.
    Using $R$=La and Eu single crystals of this system, 
    it was demonstrated in our earlier works that 
    three characteristic temperatures, \T*\ (pseudogap temperature), 
    \Tonset\ (onset temperature of fluctuating superconductivity), and \Tc\ can be defined, 
    which behave differently on the phase diagram with change of both $R$ and $x$ \cite{Okada_JPSJ,Okada_Nernst}.
    To approach the pseudogap issue further in the present study, 
    we probed the momentum dependence of the energy gap
    and compared the characteristic energy scales 
    to the above three temperatures
    focusing on the same system as in the previous study.
    All the experimental results shown in this paper consistently point to 
    the existence of three distinct energy and temperature scales 
    arising from the competition between the two states in high-\Tc\ cuprates.
    Moreover, the duality of competition and superconducting fluctuation 
    around the antinodal region is suggested to be important.
    %

\section{\label{sec:level2}Experimental}
    Single crystals of \BiR\ ($R$=La and Eu) were grown by the floating zone method \cite{Okada_crystal,Okada_Nernst}.
    The bulk sensitive ARPES spectra 
    with an ultraviolet laser (6.994 eV photons) were taken 
    by a Scienta R4000 hemispherical analyzer
    at the Institute of Solid State Physics (ISSP) \cite{Kiss}.
    %
    In this study, the total energy resolution of a photoemission 
    spectrometer ($\Delta$$E$) is defined by fitting Au spectrum 
    with the Fermi Dirac function, its intensity at each energy are broadened by Gaussian (full width at half maximum $\Delta$$E$).
    The energy relosutions of all the ARPES experiments with 6.994 eV photons shown in this paper were 
    better than 2.2 meV, and all the measurements were performed at pressures 
    below 5 $\times$ $10^{-11}$ Torr.
    %
    Prior to the ARPES measurements, 
    we carefully evaluated the doping levels of the crystals 
    with $c$-axis lattice constant, thermopower, 
    and/or inductive coupling plasma (ICP) spectroscopy.\cite{Okada_Nernst} 
    %

\begin{figure}[ltp]
\begin{center}
\includegraphics[width=0.7\columnwidth,clip]{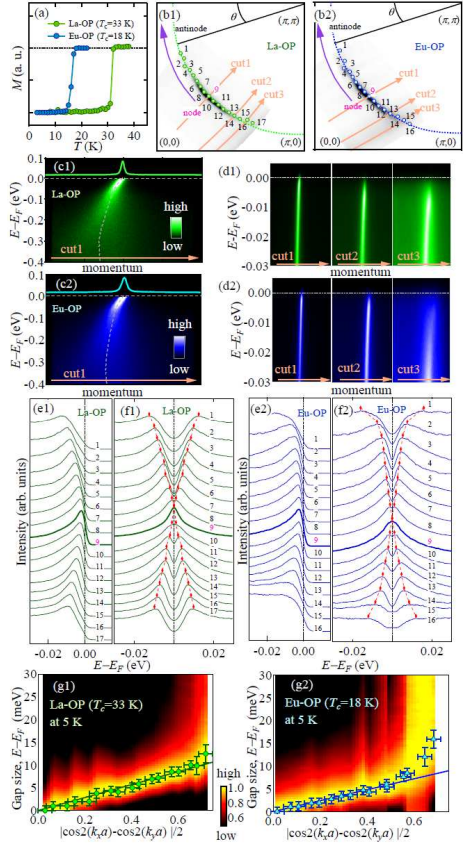}
\end{center}
  \caption{
 (Color online)
    ARPES data obtained with 6.994 eV photons at 5 K for optimally doped \BiR\ 
    with $R$=La (La-OP, \Tc=33 K) and $R$=Eu (Eu-OP, \Tc=18 K) are shown.
    (a) shows the temperature dependence of magnetization
    of the crystals.
    (b1) and (b2) show the mapping of the Fermi momentum \kF\ 
    for La-OP and Eu-OP, respectively.
    Here, the Fermi surface that was determined with 21.214 eV photons 
    in our previous study \cite{Okada_SNS} is shown with doted lines.
    (c1) and (d1) ((c2) and (d2)) show the dispersion images along 
    the momentum shown in (b1) ((b2)) for La-OP (Eu-OP).
    We show also momentum distribution curves at \EF\ in 
    the upper part of (c1) and (c2).    
    (e1) and (f1) ((e2) and (f2)) show the energy distribution curves 
    and their symmetrized spectra at \kF\ for La-OP (Eu-OP), respectively.
    (g1) ((g2)) shows the intensity map of (f1) ((f2)) together with the gap size 
    as a function of $\mid$cos($k_x$a)-cos($k_y$a)$\mid$/2 for La-OP (Eu-OP).
    The intensity is normalized to unity at the gap energy, 
    and the color scales are the same for both figures.
    }
\end{figure} 
\label{fig1.eps}

\section{\label{sec:level3}Results and Discussions}
\subsection{\label{sec:level3}$\Delta_{sc0}$: energy scale of pairing at the antinode}
    Fig. 1 shows the ARPES results obtained at 5 K with 6.994 eV photons on the optimally doped 
    \BiR\ with $R$=La (La-OP) and $R$=Eu (Eu-OP).
    As shown in Fig. 1(a), \Tc\ of the La-OP and Eu-OP samples were 33 K and 18 K, respectively.
    Figs. 1(b1) and (b2) show the Fermi momenta \kF\ 
    where the ARPES spectra were taken.
    The Fermi surfaces determined with 21.214 eV photons in our previous work 
    using samples with similar doping \cite{Okada_SNS} 
    are also shown in Figs. 1(b1) and (b2).
    Figs. 1(c1) and (d1) (Figs. 1(c2) and (d2)) show the momentum dependence of the spectral intensity of La-OP (Eu-OP) 
    along the cuts shown in Fig. 1(b1) (Fig. 1(b2)).
    The energy distribution curves at \kF\ 
    and the spectra that were symmetrized about \EF\ are shown in Figs. 1(e1)-(e2) and (f1)-(f2), respectively.
    %
    We determined the energy gap by fitting the symmetrized 
    spectra with the phenomenological spectral function,\cite{fitting_notice,Norman_fit} 
    which has been used in many other reports.\cite{Nakayama,Meng,Wei,Lee,Kanigel}
\begin{figure}[]
\begin{center}
\includegraphics[width=1\columnwidth,clip]{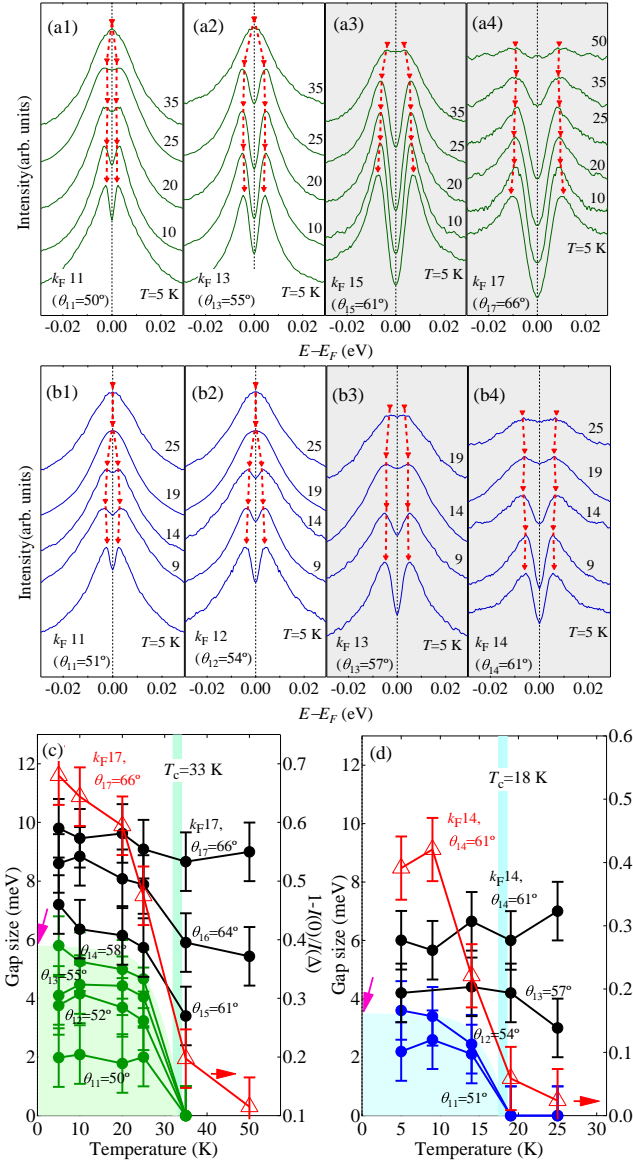}
\end{center}
\caption{
 (Color online)
    Evolution of the energy gap across \Tc\ of La-OP (\Tc=33 K) and Eu-OP (\Tc=18 K) measured with 
    laser-ARPES (6.994 eV photons).
    (a1)-(a4) and (b1)-(b4) show the temperature dependence of the  
    symmetrized ARPES spectra for La-OP and Eu-OP, respectively.
    Here, the index of \kF\ corresponds to the numbers in Figs. 1(b1) and (b2), 
    where $\theta$ is also defined.
    (c) and (d) show the temperature evolution of the gap size (left axis) 
    at various momenta across \Tc\ for La-OP and Eu-OP, respectively. 
    The hatched area shows roughly the range where 
    the energy gap was strongly temperature dependent and 
    collapsed at \Tc.
    The characteristic energy $\Delta_{sc0}^{eff}$ is indicated by the arrows on 
    the left axes of (c) and (d).
    The temperature dependence of 1-$I$(0)/$I$($\Delta$) (right axis) 
    calculated from the spectra measured at point 17 (14) for the $R$=La (Eu) sample
    is also plotted, where $I$(0) and $I$($\Delta$) are the intensity at \EF\ and at the gap edge, respectively.
    }
\label{fig:fig2.eps}
\end{figure}
\begin{figure}
\begin{center}
\includegraphics[width=0.55\columnwidth,clip]{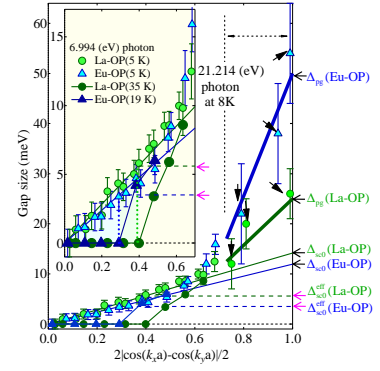}
\end{center}
\caption{
    (Color online)
    Comparison of the momentum dependence of the energy gaps 
    of La-OP (\Tc=33 K) and Eu-OP(\Tc=18 K) at $T$=5 K, which is well below \Tc, and at $T\agt$\Tc.
    The three data points closest to the antinode (indicated by arrows) are obtained using 21.214 eV      
    photons both for La-OP and Eu-OP (at 5 K with less than 20 meV resolution).\cite{Okada_SNS,notice1} 
    The three characteristic energy scales for both La-OP and Eu-OP are shown on the right axis.    
    The inset is an enlarged plot around the node to show more clearly the neighborhood of the node and $\Delta_{sc0}^{eff}$.
    }
\label{fig:fig3.eps}
\end{figure} 
\begin{figure*}
\begin{center}
\includegraphics[width=2\columnwidth,clip]{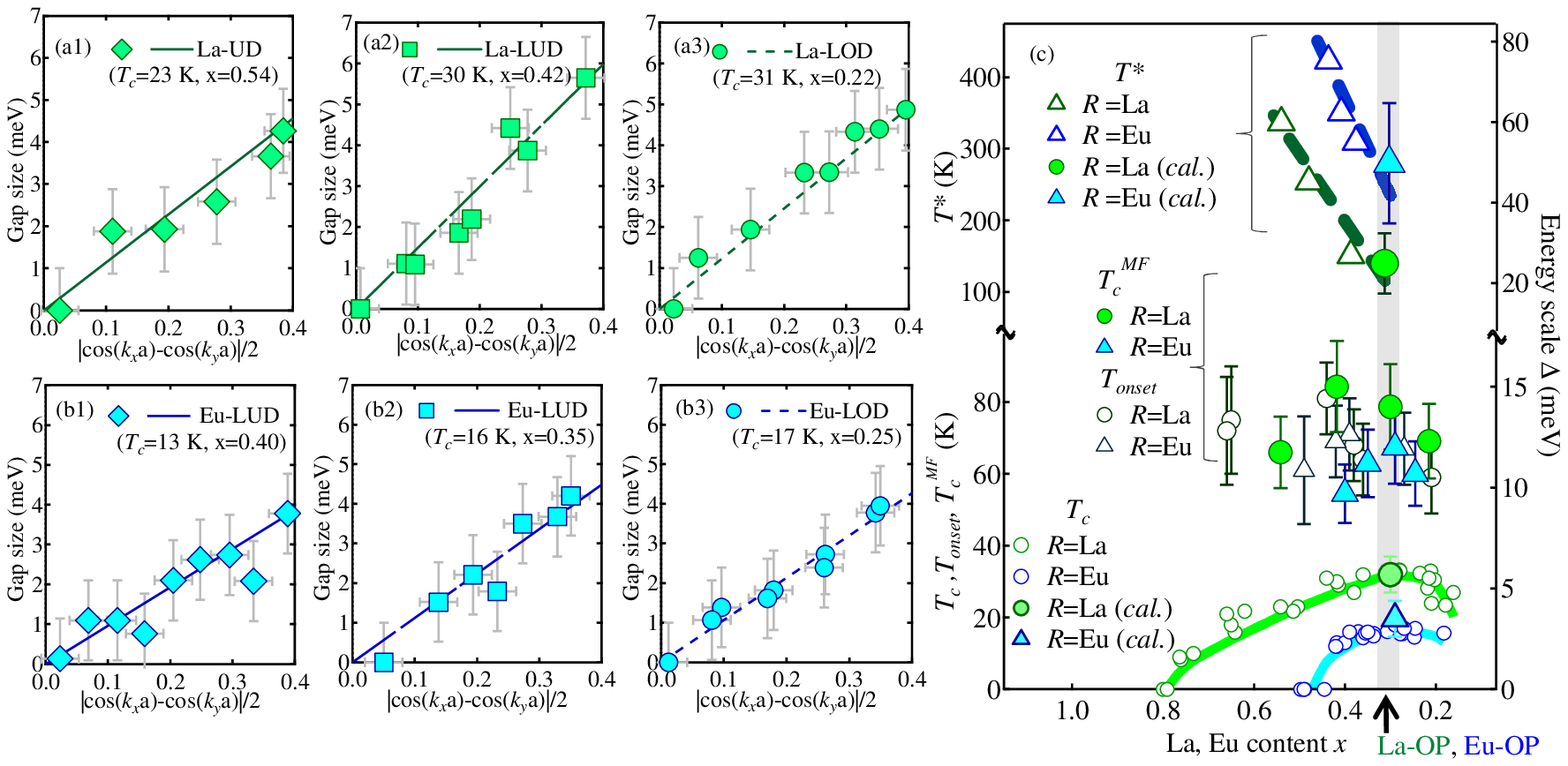}
\end{center}
\caption{
 (Color online)
    Energy gaps around the node 
    of \BiR\ with $R$=La and Eu samples determined by 
    the laser-ARPES measurements (6.994 eV photons) 
are shown in (a1)-(a3) and (b1)-(b3), respectively.
    All data shown here were measured at $T\leq$5 K, which is well below \Tc.
    (c) Phase diagram of the three characteristic temperatures.
    The data of \T*, \TMF, and \Tc\ plotted with solid symbols 
    were calculated from $\Delta_{pg}$, $\Delta_{sc0}$, and $\Delta_{sc0}^{eff}$ 
    assuming 2$\Delta$/$k_{B}$$T$=4.3.     
    The right axis of (c) 
    gives the energy scale $\Delta$ that is connected to 
    the temperature scale (left axis) by the above relation. 
    The \Tc, \Tonset, and \T*\ data shown with empty symbols 
    are from our previous studies \cite{Okada_Nernst,Okada_JPSJ,Okada_crystal}.
        }
\label{fig:fig4.eps}
\end{figure*} 

    The gap size with La-OP and Eu-OP are plotted as a function of 
    $\mid$cos($k_x$a)-cos($k_y$a)$\mid$/2 in Figs.\ 1(g1) and (g2), respectively.
    Since a $d$-wave gap is expressed as 
    $\Delta$=$\Delta_{sc0}$$\mid$cos($k_x$a)-cos($k_y$a)$\mid$/2, Figs.\ 1(g1) and (g2) show
    that the gap has a pure $d$-wave form around the node for both La-OP and Eu-OP.
    On the other hand, the data points deviated from the $d$-wave form near the antinode.
    This deviation is accompanied by a huge broadening of 
    the spectral linewidth as is evident from the image plot of the ARPES spectra shown in the same figure. 
    We determined $\Delta_{sc0}$ by fitting the linear part of the data,
    which gave 14.1 meV and 12.0 meV for La-OP (\Tc=33 K) and Eu-OP (\Tc=18 K), respectively.
    Hence, the value of $\Delta_{sc0}$ changed together with \Tc.
    However, the difference in $\Delta_{sc0}$ is not as large as the change of \Tc\ 
    since the ratio of $\Delta_{sc0}$ (14.1/12.0$\approx$1.23) 
    is much smaller than the \Tc\ ratio 33/18$\approx$1.8.
    This is in strong contrast to conventional superconductors, 
    for which \Tc\ scales with the binding energy of the paired electrons $\Delta_{sc0}$, 
    and suggests the possible existence of an energy scale other than $\Delta_{sc0}$ corresponding to \Tc.

    \subsection{\label{sec:level3}$\Delta_{sc0}^{eff}$: energy scale related to \Tc }
    The temperature evolution of the symmetrized spectrum of La-OP and Eu-OP across \Tc\ 
    is shown in Figs. 2(a1)-(a4) and 2(b1)-(b4) for various momenta.
    Figs. 2(c) and 2(d) show the temperature dependence of the gap size (left axis) 
    together with the gap depth (right axis, see the caption for the definition) for the momentum that is closest to the antinode 
    among the data shown in Fig. 2.
    From this figure, a sudden change in the gap depth was observed across \Tc\ 
    although the gap size for this momentum did not show obvious change.
    On the contrary, a more dramatic change happened across \Tc\ at momenta around the node: 
    the energy gaps all collapsed simultaneously at \Tc.
    %
    Based on this experimental observation, 
    we can define a characteristic energy scale $\Delta_{sc0}^{eff}$, 
    which is indicated by the arrows in Figs. 2(c) and (d) on the left axes.
    %
    When the energy gap at $T$=0 was smaller than this characteristic energy $\Delta_{sc0}^{eff}$,    
    it decreased abruptly to zero at \Tc,
    while the energy gap remained finite above \Tc\ if it was larger 
    than $\Delta_{sc0}^{eff}$ at $T$=0.
    The existence of such a characteristic energy is consistent with other recent reports.\cite{Oda,Lee,Kanigel_PRL_no2,Nakayama,Kurosawa}
    %
    Here, $\Delta_{sc0}^{eff}$ of La-OP(\Tc=33 K) and Eu-OP(\Tc=18 K) are 
    5.6 $\pm$ 1.1 meV and 3.5 $\pm$ 1.1 meV, respectively.
    The ratio of $\Delta_{sc0}^{eff}$ between the two samples are about 1.6 $\pm$ 0.5, 
    which is close to the ratio of \Tc\ ($\approx$1.8) within experimental error.
    Moreover, the values of $2\Delta_{sc0}^{eff}$/$k_{B}T_{c}$ for La-OP (3.9 $\pm$ 0.8) 
    and Eu-OP (4.5 $\pm$ 1.4) were close to the value observed 
    by Andreev reflection experiments on a wide range of cuprates with various \Tc s.\cite{Andreef}
    %
    These quantitative comparisons indicate 
    that $\Delta_{sc0}^{eff}$ can be attributed to the energy scale corresponding to \Tc.
 
    \subsection{\label{sec:level3}$\Delta_{pg}$: energy scale related to a competing pseudogap state}
    The question to be addressed next is why $\Delta_{sc0}^{eff}$ is much lower than $\Delta_{sc0}$.
    Figure 3 shows the momentum dependence of the energy gap of the La-OP and Eu-OP samples
    at $T$=5 K and $T\agt$ \Tc.
    The energy gap in the antinodal region
    obtained with 21.214 eV photons (at 5 K with less than 20 meV resolution) 
    in our previous studies \cite{Okada_SNS,notice1} are also included.
    In contrast to the gap around the node, 
    the antinodal gap $\Delta_{pg}$ is clearly larger for Eu-OP (\Tc = 18 K) 
    than La-OP (\Tc = 33 K) 
    showing that the nodal and antinodal gaps depend differently on \Tc, 
    which is qualitatively consistent with STM/STS results\cite{Sugimoto}.
    %
    Previously, we observed that 
    the coherent part of the remnant Fermi surface, 
    where clear peaks were observed in the ARPES spectra at the superconducting state,  
    narrowed with increasing $\Delta_{pg}$.\cite{Okada_SNS}
    This observation suggested that $\Delta_{pg}$ shrinks the coherent part of 
    the remnant Fermi surface, which naturally decreases the superfluid density. 
    All our experimental observations suggest that the antinodal pseudogap 
    state characterized by $\Delta_{pg}$ competes with superconductivity and 
    supresses \Tc, resulting in the deviation of $\Delta_{sc0}^{eff}$ from $\Delta_{sc0}$.
    The $T$=5 K data of fig. 3 is the indication of the co-existence of the 
    competing state with superconductivity.
    This supports the existence of two different momentum 
    dependent gap components \cite{Tanaka,Kondo_PRL,Terashima,Lee, Vidya, Kondo_Nature,He,Hashimoto_Bi2201,Yoshida}; 
    that is the antinodal gap $\Delta_{pg}$ has its origin in a competing state 
    with no direct relation to the $d$-wave superconductivity\cite{Tallon}.

    \subsection{\label{sec:level3}The intimate relation between $\Delta_{sc0}$ and \Tonset}
    Figs. 4(a1)-(a3) and (b1)-(b3) show the momentum dependence of the energy gap around the node 
    of various \BiR\ crystals with different $x$ for $R$=La and Eu, respectively. 
    The results indicate that the slope of the energy gap as a function of 
    $\mid$cos($k_x$a)-cos($k_y$a)$\mid$/2 did not change much with 
    $R$ or $x$ despite the large variation of \Tc.
    This relatively insensitive behaviour of $\Delta_{sc0}$ mimics that of \Tonset, 
    the temperature below which the Nernst signal starts to 
    be enhanced with decreasing temperature \cite{Okada_Nernst}.
    To address this similarity more quantitatively, 
    we calculated the mean field transition temperature (\TMF) 
    based on the weak coupling theory of $d$-wave superconductivity (2$\Delta_{sc0}$/$k_B$\TMF=4.3).
    Fig. 4(c) shows \TMF\ calculated from the data shown in Figs. 4(a1)-(a3) and (b1)-(b3)
    together with \Tc, \Tonset, and \T*\ reported 
    in our previous studies \cite{Okada_Nernst,Okada_JPSJ,Okada_crystal}.
    Interestingly, we found that \TMF\ agrees quite well with \Tonset.
    We think this agreement implies that the energy scale $\Delta_{sc0}$ 
    is related to the onset pairing temperature \Tonset.
    The large difference between \Tonset\ and \Tc\ ($\Delta_{sc0}$ and $\Delta_{sc0}^{eff}$) indicates that there exists a large superconducting fluctuation.
    The phenomenological explanation of the existence of 
    a large superconducting fluctuation is due to 
    weak perturbation of the pairing energy $\Delta_{sc0}$ 
    by stabilization of the competing state (increasing $\Delta_{pg}$).
    %
    We think that this is consistent with the existence of a 
    relatively homogeneous gap despite the large variation of the pseudogap 
    in real space as was revealed by recent STM experiments.\cite{Pushp,Mike}

    %
    Note that while $\Delta_{sc0}$ is the pairing energy scale at the antinode, 
    the energy gap observed at this momentum is not $\Delta_{sc0}$ but $\Delta_{pg}$.
    As shown in fig. 4(c), 
    the characteristic temperature scale $\Delta_{sc0}$ is related to \Tonset. 
    Therefore, the observed relation between \Tonset\ and $\Delta_{sc0}$ suggests the existence of both 
    competition and superconducting fluctuation at momenta around 
    the antinode below \Tonset.
    %
    
    \subsection{\label{sec:level3}Three energy- and temperature- scales in high-\Tc\ cuprates}
    %
    In fig. 4(c), we plot all the experimentally obtained energy and temperature scales 
    with changing both $x$ and $R$ in \BiR.
    This phase diagram clearly shows the existence of 
    three energy ($\Delta_{sc0}^{eff}$, $\Delta_{sc0}$, and $\Delta_{pg}$) 
    and temperature (\Tc, \Tonset, and \T*) scales connected by the relation 2$\Delta$/$k_{B}$$T$=4.3.
    The natural consistent picture led by the phase diagram of fig. 4 (c) is that 
    the pseudogap state (characterized by $\Delta_{pg}$ and \T*) 
    suppresses coherent superconductivity ($\Delta_{sc0}^{eff}$ and \Tc) 
    while keeping the pairing strength ($\Delta_{sc0}$ and \Tonset) similar\cite{notice_momentum}.
    %
    Therefore, the competing state kills superconductivity mainly by enhancing fluctuation of 
    superconducting order through reducing superfluid density (phase stiffness).
    In other words, one may say that the $competition$ enhances the 
    $superconducting$ $fluctuation$.
    We think the conclusion in this paper can be extended more or less to 
    all the high-\Tc\ cuprates
    including systems that have a comparable $\Delta_{sc0}$ and $\Delta_{pg}$, 
    such as Bi$_2$Sr$_2$CaCu$_2$O$_y$.
    %

    \section{\label{sec:level4}Summary}
    In summary, we compared the momentum dependence of the gap function and 
    the characteristic temperature scales of \BiR\ ($R$=La and Eu).
    All the experimental results point towards the existence of 
    three distinct energy and temperature scales 
    corresponding to the competing pseudogap state, and the incoherent 
    and coherent superconducting states.
    Accounting for all these three phenomena consistently would be 
    crucial for understanding the pseudogap issue in high-\Tc\ cuprates.

    \section{\label{sec:level5}Acknowledgement}
    We thank E. Hudson and V. Madhavan for useful discussions.
    We also thank Y. Hamaya and S. Arita for experimental assistance.
    Y. O. thanks JSPS for a financial support.


\begin{thebibliography}{99}
\bibitem{Timsk}
    T. Timusk and B. Statt, 
    Rep. Prog. Phys. {\bf 62}, 61 (1999).
\bibitem{Norman}
        M. R. Norman, D. Pines, and C. Kallin, 
        Advances in Physics, {\bf 54}, 715 (2005). 
\bibitem{Huffner}
        S. Huffner, M. A. Hossain, A. Damascelli, and G. A. Sawatzky, 
        Rep. Prog. Phys. {\bf 71}, 062501 (2008).
\bibitem{Emery}
    V. J. Emery and S. A. Kivelson,
    Nature {\bf 374}, 434 (1995).
\bibitem{Varma}
    C. M. Varma, 
    Phys. Rev. B {\bf 55}, 14554 (1997).
\bibitem{Chacravati}
        S. Chakravarty, R. B. Laughlin, D. K. Morr, and Chetan Nayak,
        Phys. Rev. B {\bf 63}, 094503 (2001).
\bibitem{Kanigel}
        A. Kanigel, M. R. Norman, M. Randeria, U. Chatterjee, S. Souma, A. Kaminski, 
        H. M. Fretwell, S. Rosenkranz, M. Shi, T. Sato, T. Takahashi, Z. Z. Li, H. Raffy, K. Kadowaki,
        D. Hinks, L. Ozyuzer, and J. C. Campuzano, 
        Nature Phys. {\bf 2}, 447 (2006).
\bibitem{Valla}
    T. Valla, A. V. Fedorov, Jinho Lee, J. C. Davis, and G. D. Gu, 
    Science {\bf 314}, 1914 (2006).
\bibitem{Wei}
    J. Wei, Y. Zhang, H. W. Ou, B. P. Xie, 
    D. W. Shen, J. F. Zhao, L. X. Yang, M. Arita, 
    K. Shimada, H. Namatame, M. Taniguchi, Y. Yoshida, H. Eisaki, and D. L. Feng, 
    Phys. Rev. Lett. {\bf 101}, 097005 (2008).
\bibitem{Meng}
    J. Meng, W. Zhang, G. Liu, L. Zhao, H. Liu, X. Jia, W. Lu, 
    X. Dong, G. Wang, H. Zhang, Y. Zhou, Y. Zhu, X. Wang, 
    Z. Zhao, Z. Xu, C. Chen, and X. J. Zhou, 
    Phys. Rev. B {\bf 79}, 024514 (2009).
\bibitem{Nakayama}
    K. Nakayama, T. Sato, Y. Sekiba, K. Terashima, 
    P. Richard, T. Takahashi, K. Kudo, N. Okumura, T. Sasaki, and N. Kobayashi,  
    Phys. Rev. Lett. {\bf 102}, 227006 (2009).
\bibitem{Shi}
    M. Shi, J. Chang, S. Pailhes, M. R. Norman, J. C. Campuzano, 
    M. Mansson, T. Claesson, O. Tjernberg, A. Bendounan, L. Patthey, 
    N. Momono, M. Oda, M. Ido, C. Mudry, and J. Mesot, 
    Phys. Rev. Lett. {\bf 101}, 047002 (2008).
\bibitem{Corson}
        J. Corson, R. Mallozzi, J. Orenstein, 
        J. N. Eckstein, and I. Bozovic, 
        Nature {\bf 398}, 221 (1999).
\bibitem{Xu}
        Z. A. Xu, N. P. Ong, Y. Wang, T. Kakeshita, and S. Uchida, 
        Nature {\bf 406}, 486 (2000).
\bibitem{Wang_Nernst_diamagnetism}
        Y. Wang, L. Li, M. J. Naughton, G. D. Gu, S. Uchida, and N. P. Ong,
        Phys. Rev. Lett. {\bf 95}, 247002 (2005).
\bibitem{Lee_STM}
    J. Lee, K. Fujita, A. R. Schmidt, C. K. Kim, H. Eisaki, 
    S. Uchida, and J. C. Davis,
    Science {\bf 325}, 1099 (2009).
\bibitem{Tanaka}
        K. Tanaka, W. S. Lee, D. H. Lu, A. Fujimori, T. Fujii, 
        Risdiana, I. Terasaki, D. J. Scalapino, T.   P. Devereaux, 
        Z. Hussain, and Z. X. Shen, 
        Science {\bf 314}, 1910 (2006).
\bibitem{Kondo_PRL}
        T. Kondo, T. Takeuchi, A. Kaminski, S. Tsuda, and S. Shin, 
        Phys. Rev. Lett. {\bf 98}, 267004 (2007).
\bibitem{Terashima}
        K. Terashima, H. Matsui, T. Sato, T. Takahashi, 
        M. Kofu, and K. Hirota,  
        Phys. Rev. Lett. {\bf 99}, 017003 (2007).  
\bibitem{Lee}
       W. S. Lee, I. M. Vishik, K. Tanaka, D. H. Lu, T. Sasagawa, 
       N. Nagaosa, T. P. Devereaux, Z. Hussain, and  Z.-X. Shen, 
       Nature {\bf 450}, 81 (2007).  
\bibitem{Vidya}
       J.-H. Ma, Z.-H. Pan, F. C. Niestemski, M. Neupane, Y.-M. Xu, P. Richard, 
        K. Nakayama, T. Sato, T. Takahashi, H.-Q. Luo, L. Fang, H.-H. Wen, Ziqiang Wang, 
        H. Ding, and V. Madhavan,
        Phys. Rev. Lett. {\bf 101}, 207002 (2008).
\bibitem{Kondo_Nature}
       T. Kondo, R. Khasanov, T. Takeuchi, J. Schmalian, and A. Kaminski,
        Nature {\bf 457}, 296 (2008).
\bibitem{Kondo_condmat}
        T. Kondo, Y. Hamaya, A. D. Palczewski, T. Takeuchi, 
        J. S. Wen, Z. J. Xu, G. Gu, J. Schmalian, A. Kaminski, 
        arXiv:1005.5309
\bibitem{He}
       R. H. He, K. Tanaka, Sung-Kwan Mo, T. Sasagawa, 
       M. Fujita, T. Adachi, N. Mannella, K. Yamada, 
       Y. Koike, Z. Hussain, and Z. X. Shen, 
       Nat. Phys. {\bf 5}, 119 (2008).
\bibitem{Hashimoto_Bi2201}
        M. Hashimoto, T. Yoshida, A. Fujimori, D.H. Lu, 
        Z.-X. Shen, M. Kubota, K. Ono, 
        M. Ishikado, K. Fujita, and S. Uchida, 
        Phys. Rev. B {\bf 79}, 144517 (2009).
\bibitem{Yoshida}
    T. Yoshida, M. Hashimoto, S. Ideta, A. Fujimori, 
    K. Tanaka, N. Mannella, Z. Hussain, Z.-X. Shen, 
    M. Kubota, K. Ono, S. Komiya, Y. Ando, H. Eisaki, and S. Uchida, 
    Phys. Rev. Lett. {\bf 103}, 037004 (2009).
\bibitem{Doug}
        W. D. Wise, M. C. Boyer, K. Chatterjee, 
        T. Kondo, T. Takeuchi, H. Ikuta, Yayu Wang, and E. W. Hudson 
        Nat. Phys. {\bf 4}, 696 (2008).
\bibitem{Hashimoto_Nat_Phys}
    M. Hashimoto, R. H. He, K. Tanaka, J. P. Testaud, W. Meevasana, 
    R. G. Moore, D. Lu, H. Yao, Y. Yoshida, H. Eisaki, T. P. Devereaux, 
    Z. Hussain, and Z. X. Shen,
    Nat. Phys. {\bf 6}, 414 (2010).
\bibitem{Nameki}
       H. Nameki, M. Kikuchi, and Y. Syono, 
        Physica C {\bf 234}, 255 (1994).
\bibitem{Eisaki}
        H. Eisaki, N. Kaneko, D. L. Feng, A. Damascelli, 
        P. K. Mang, K. M. Shen, Z.-X. Shen, and M. Greven, 
        Phys. Rev. B {\bf 69}, 064512 (2004).
\bibitem{Fujita}
        K. Fujita, T. Noda, K. M. Kojima, 
        H. Eisaki, and S. Uchida, 
        Phys. Rev. Lett. {\bf 95}, 097006 (2005).
\bibitem{Okada_Nernst}
    Y. Okada, Y. Kuzuya, T. Kawaguchi, and H. Ikuta, 
    Phys. Rev. B {\bf 81}, 214520 (2010).
\bibitem{Okada_JPSJ}
        Y. Okada, T. Takeuchi, T. Baba, S. Shin, and H. Ikuta,
        J. Phys. Soc. Jpn. {\bf 77}, 074714 (2008).
\bibitem{Okada_crystal}
        Y. Okada and H. Ikuta, 
        Physica C {\bf 445-448}, 84 (2006).
\bibitem{Kiss}
    T. Kiss, T. Shimojima, K. Ishizaka, A. Chainani, 
    T. Togashi, T. Kanai, X. Y. Wang , C. T. Chen, 
    S. Watanabe, and S. Shin, 
    Rev. Sci. Instrum. {\bf 79}, 023106 (2008).
\bibitem{Sugimoto}
        A. Sugimoto, S. Kashiwaya, H. Eisaki, H. Kashiwaya, 
        H. Tsuchiura, Y. Tanaka, K. Fujita, and S. Uchida, 
        Phys. Rev. B {\bf 74} (2006) 094503.
\bibitem{Okada_SNS}
       Y. Okada, T. Takeuchi, A. Shimoyamada, 
        S. Shin, and H. Ikuta, 
        J. Phys. Chem. Solids {\bf 69}, 2989 (2008).
\bibitem{fitting_notice}
     In this study, we fit the symmetrized spectra with the spectral function 
    $A$($k_F$,$\omega$)=1/$\pi$$\cdot$Im($\Sigma$)/{($\omega$-Re($\Sigma$))$^2$+(Im($\Sigma$))$^2$} 
    using a simple phenomenological self energy 
    $\Sigma$($k_F$,$\omega$)=$-$$i\Gamma_1$$+$$\Delta^2$$/$($\omega^2$$+$$i\Gamma_0$), 
    where $\Delta$ is gap size, $\Gamma_1$ is single particle scattering rate, 
    and $\Gamma_0$ is an inverse pair lifetime as shown in ref. 39. 
\bibitem{Norman_fit}
    M. R. Norman, M. Randeria, H. Ding, and J. C. Campuzuano, 
    Phys. Rev. B. {\bf 57}, R11093 (1998).
\bibitem{Oda}
    M. Oda, R. M. Dipasupil, N. Momono, and M. Ido, 
    J. Phys. Soc. Jpn. {\bf 69}, 983 (2000).
\bibitem{Kurosawa}
    T. Kurosawa, T. Yoneyama, Y. Takano, M. Hagiwara, 
    R. Inoue, N. Hagiwara, K. Kurusu, K. Takeyama, N. Momono, M. Oda, and M. Ido,
     Phys. Rev. B {\bf 81}, 094519 (2010).
\bibitem{Kanigel_PRL_no2}
    A. Kanigel, U. Chatterjee, M. Randeria, M. R. Norman, S. Souma, M. Shi, 
    Z. Z. Li, H. Raffy, and J. C. Campuzano, 
    Phys. Rev. Lett. {\bf 99}, 157001 (2007).
\bibitem{Andreef}
    G. Deutscher, Nature {\bf 397}, 410 (1998), and references therein.
\bibitem{notice1}
    Probing the electronic structure around the antinode with 6.994 eV photons 
    was difficult due to the matrix element effect. 
    %
    However, there was a smooth connection between the gap sizes 
    obtained with 6.994 eV and 21.214 eV photons.
    %
    Also, the deviation from the $d$-wave gap for Eu-OP was stronger 
    than La-OP in both measurements.
    %
    Hence, the data with different photon energies are consistent with each other.
    %
    \bibitem{Mike}
        M. C. Boyer, W. D. Wise, Kamalesh Chatterjee, Ming Yi, 
        T. Kondo, T. Takeuchi, H. Ikuta, E. W. Hudson, 
        Nat. Phys. {\bf 3}, 802 (2007).
    %
    \bibitem{Pushp}
        A. Pushp, C. V. Parker, A. N. Pasupathy, K. K. Gomes, 
        S. Ono, J. Wen, Z. Xu, G. Gu, and A. Yazdani, 
        Science {\bf 324}, 1689 (2009).
    \bibitem{Tallon}
        J. L. Tallon and J. W. Loram, 
        Physica C {\bf 349}, 53 (2000), and references therein.
        \bibitem{notice_momentum}
        As shown in this paper, ARPES directly gives us the important three energy scales 
        characterized around the node (probably rather homogeneous) 
        and antinode (probably inhomogeneous) leading to a deeper understanding of the pseudogap issue in high-\Tc\ cuprates.
        %
        Most of our discussions and conclusions in this paper are drawn in the momentum space, however, 
        we think our conclusion can be reconciled with the real space picture of an inhomogeneous electronic structure.
        %
        Especially, 
        the origin of the duality between competition and superconducting fluctuation around the antinode
        is possibly related to the inhomogeneous quasiparticle excitations in real space.
\end{thebibliography}
\end{document}